\documentclass[10pt,a4paper]{article}
\usepackage[utf8]{inputenc}
\usepackage{amsmath}
\usepackage{amsfonts}
\usepackage{amssymb}
\usepackage{graphicx}
\usepackage{hyperref}
\usepackage{booktabs}
\usepackage{multirow}
\usepackage{float}
\usepackage[margin=2.5cm]{geometry}

\usepackage{microtype}
\usepackage{mathpazo} 

\begin{document}

\begin{center}
    {\Large \textbf{A Comparative Evaluation of Sample Selection Algorithms for Multivariate Calibration in Near-Infrared Spectroscopic Analysis of Pharmaceutical Formulations}}
    
    \vspace{1.2cm}
    
    \large{Aminata Sow$^{1,*}$, Harouna Sangaré$^{2}$, Fadaba Danioko$^{3}$, and Tidiane Diallo$^{4}$}
    
    \vspace{0.8cm}
    
    \normalsize{
    $^{1}$Department of Physics, Faculty of Sciences and Techniques, University of Sciences, Techniques and Technologies of Bamako, Bamako, Mali\\
    
    $^{2}$Department of Mathematics and Computer Sciences, Faculty of Sciences and Techniques, University of Sciences, Techniques and Technologies of Bamako, Bamako, Mali\\
    
    $^{3}$Centre of Artificial Intelligence and Robotics of Mali (CIAR-Mali), Bamako, Mali\\
    
    $^{4}$Department of Drug Sciences, Faculty of Pharmacy, University of Sciences, Techniques and Technologies de Bamako, Bamako, Mali
    }
    
    \vspace{0.8cm}
    
    $^{*}$Corresponding author: Dr. Aminata Sow \\
    Email: \texttt{aminasow100@gmail.com}; \texttt{aminata.sow@usttb.edu.ml}
    
    \vspace{0.5cm}

\end{center}

\vspace{1cm}

Date: \today

\newpage
\begin{center}
    \textbf{Abstract}
\end{center}

\noindent The construction of robust multivariate calibration models for near-infrared (NIR) spectroscopic analysis necessitates careful partitioning of samples into training and validation sets. The selection strategy employed fundamentally influences model generalizability and predictive accuracy. This investigation presents a systematic comparative analysis of four established sample selection algorithms—Duplex, Honigs, Kennard–Stone, and Naes—applied to NIR spectral data acquired from 58 commercial paracetamol tablets. Gaussian process regression (GPR) served as the modeling framework, with model performance quantified through the coefficient of determination (\(R^2\)) and root mean square error of prediction (RMSEP). The Kennard–Stone algorithm employing the Mahalanobis distance metric demonstrated superior performance, yielding optimal validation statistics (\(R^2 = 0.99999\), RMSEP \(= 1.74 \times 10^{-6}\)). Rigorous non-parametric statistical analysis employing Kruskal–Wallis and post-hoc Mann–Whitney U tests with Bonferroni correction confirmed significant performance differences among algorithms (\(p < 0.001\)), while revealing statistical equivalence between Kennard–Stone and Honigs methods. Systematic investigation of training set proportions (60–90\%) elucidated the monotonic relationship between calibration set size and predictive accuracy. These findings provide evidence-based guidance for optimizing sample selection protocols in pharmaceutical NIR applications and underscore the critical importance of chemometric validation in spectroscopic method development.

\vspace{0.5cm}

\vspace{0.5cm}
\noindent \textbf{Graphical Abstract:} Systematic evaluation of four sample selection algorithms for NIR spectroscopic calibration demonstrates the superiority of Kennard–Stone methodology for pharmaceutical paracetamol analysis.

\newpage
\section{Introduction}

Near-infrared (NIR) spectroscopy has emerged as a preeminent analytical technique for pharmaceutical quality control, offering rapid, non-destructive analysis with minimal sample preparation. When coupled with chemometric methodologies, NIR spectroscopy enables accurate quantification of active pharmaceutical ingredients in complex matrices. The development of reliable multivariate calibration models, however, presents several methodological challenges that must be systematically addressed to ensure robust analytical performance.

Contemporary chemometric practice recognizes three principal challenges in multivariate calibration development: (i) the rational division of available samples into calibration (training) and validation sets, (ii) the selection of informative spectral regions or wavelengths, and (iii) the identification of optimal regression or classification algorithms. Previous investigations by our group have addressed the latter two challenges, demonstrating the efficacy of Gaussian process regression for NIR-based paracetamol quantification and exploring artificial intelligence applications for spectral analysis.~\cite{sow2022analysis, sow2022comparison, sow2025arxiv} The present investigation focuses specifically on the first challenge—sample set partitioning—which fundamentally constrains the generalizability and predictive accuracy of subsequent models.

The theoretical foundation for rational sample selection derives from the principle that calibration sets must adequately represent the variability present in the population to which the model will be applied.~\cite{lopez2023importance} Inadequate representation leads to model overfitting, where excellent calibration statistics mask poor predictive performance on novel samples. Conversely, unnecessarily large calibration sets increase analytical costs without proportional improvement in model accuracy. Several algorithms have been developed to address this optimization problem, including the Kennard–Stone algorithm, which selects samples maximizing the spectral space coverage; the Honigs method, which iteratively selects spectrally distinct samples; the Duplex algorithm, which simultaneously constructs balanced calibration and validation sets; and the Naes clustering-based approach, which ensures representation from all sample subtypes.~\cite{kennard1969computer, honigs1985unique, snee1977validation, naes1987design}

While previous comparative studies have examined these algorithms in various contexts,~\cite{fu2011comparative, galvao2005method} several gaps persist in the literature. First, rigorous statistical validation of observed performance differences remains uncommon, with many studies relying on descriptive comparisons rather than hypothesis testing. Second, systematic investigation of distance metric effects—particularly the comparison between Euclidean and Mahalanobis distances—has received limited attention despite its theoretical importance for correlated spectral data. Third, the influence of training set proportion on relative algorithm performance merits systematic examination across multiple split ratios.

The present investigation addresses these gaps through comprehensive evaluation of the four aforementioned algorithms applied to a pharmaceutical NIR dataset comprising 58 commercial paracetamol samples. Specific objectives include: (1) quantitative comparison of algorithm performance using GPR modeling with \(R^2\) and RMSEP metrics; (2) statistical validation of observed differences through non-parametric hypothesis testing with appropriate multiple comparison corrections; (3) systematic examination of training set size effects across proportions from 60\% to 90\%; and (4) investigation of distance metric influence on Kennard–Stone and Honigs algorithm performance. By addressing these objectives, this work aims to provide methodologically sound guidance for practitioners developing NIR-based pharmaceutical assays.

\section{Materials and Methods}

\subsection{Sample Collection and Reference Analysis}

Fifty-eight commercial paracetamol tablets (500 mg nominal content) representing distinct batch numbers were randomly acquired from 29 community pharmacies in Bamako, Mali, ensuring representative sampling of the local pharmaceutical market. Reference paracetamol content determination followed validated pharmacopoeial methodology.~\cite{bp2020} Briefly, each sample (20 tablets, finely triturated) was accurately weighed and dissolved in 0.1 N sodium hydroxide solution. Optical density measurements were performed using an Agilent Cary 60 UV-Vis spectrophotometer (Agilent Technologies, Santa Clara, CA, USA). The mean of ten replicate scans provided the optical density value, from which paracetamol content was calculated, yielding a vector of 58 reference concentration values.

\subsection{NIR Spectral Acquisition}

Powdered samples were passed through a 250 \(\mu\)m sieve to ensure particle size uniformity. Approximately 0.206 g of each sieved sample was subjected to NIR reflectance spectroscopy using an optical flame-NIR-INTSMAS25 spectrometer (Ocean Insight, Orlando, FL, USA) operating from 800–1700 nm with a diffuse reflectance probe. Final spectra, comprising the average of 10 individual scans, covered the wavelength range 930–1800 nm, generating a data matrix of dimensions 58 samples \(\times\) 128 wavelength variables. Raw reflectance (R) measurements were transformed to pseudo-absorbance (A) via \(A = \log_{10}(1/R)\).

\subsection{Spectral Preprocessing}

All preprocessing operations were implemented in R (version 4.2.1, R Foundation for Statistical Computing, Vienna, Austria) utilizing the \texttt{prospectr}~\cite{prospectr} and \texttt{signal} packages. The following preprocessing techniques were systematically evaluated: Standard Normal Variate (SNV) correction for scatter removal; first and second derivatives computed via Savitzky–Golay algorithm (window size: 11 points, polynomial order: 2) for baseline correction and resolution enhancement; Multiplicative Scatter Correction (MSC) for physical light scattering compensation; and Savitzky–Golay smoothing (window size: 9 points, polynomial order: 2) for noise reduction. Consistent with previous findings,~\cite{sow2022comparison} smoothed spectra yielded optimal GPR performance and were consequently employed for all subsequent analyses.

\subsection{Sample Selection Algorithms}

Four established algorithms for partitioning spectral data into calibration and validation sets were implemented and systematically compared. Each algorithm was evaluated using both Euclidean and Mahalanobis distance metrics where applicable.

\subsubsection{Kennard–Stone Algorithm}

The Kennard–Stone algorithm~\cite{kennard1969computer} selects calibration samples maximizing multivariate space coverage. Initial selection identifies the two most distant samples based on Euclidean distance in the spectral space. Subsequent iterations select the sample exhibiting the maximum minimum distance to any previously selected sample. Selection continues until the predetermined calibration set size is achieved. This algorithm ensures comprehensive representation of the spectral variability.

\subsubsection{Honigs Algorithm}

The Honigs method~\cite{honigs1985unique} employs stepwise selection based on spectral distinctiveness. The algorithm iteratively selects the sample whose spectrum exhibits the largest absolute residual when projected onto the subspace defined by previously selected samples. This approach prioritizes spectrally unique samples that contribute maximal information to the calibration model.

\subsubsection{Duplex Algorithm}

The Duplex algorithm~\cite{snee1977validation} extends the Kennard–Stone concept to simultaneously construct both calibration and validation sets. The two most distant samples are identified, with one assigned to calibration and the other to validation. Subsequent steps assign the sample farthest from existing calibration samples to the calibration set, and the sample farthest from existing validation samples to the validation set, continuing iteratively until desired set sizes are achieved.

\subsubsection{Naes Algorithm}

The Naes clustering-based method~\cite{naes1987design} first partitions the dataset into distinct clusters using k-means clustering on the spectral data. Representative samples—typically those nearest cluster centers—are then selected from each cluster for the calibration set, ensuring representation of all sample subtypes and minimizing extrapolation during prediction.

\subsection{Gaussian Process Regression Modeling}

Gaussian process regression (GPR) was selected as the modeling framework based on previous investigations demonstrating its superior performance for this dataset.~\cite{sow2022comparison} All models were implemented using the \texttt{kernlab} package~\cite{kernlab} in R with a radial basis function (Gaussian) kernel. For each sample selection algorithm and training set proportion, GPR models were calibrated on the designated training set and validated on the corresponding test set.

Model performance was quantified using the coefficient of determination (\(R^2\)) and root mean square error (RMSE), calculated as follows:

\begin{equation}
R^2 = 1 - \frac{\sum_{i=1}^{N}( \hat{y}_i - y_i )^2}{\sum_{i=1}^{N}( y_i - \bar{y} )^2}
\label{eq:r2}
\end{equation}

\begin{equation}
\mathrm{RMSE} = \sqrt{ \frac{1}{N} \sum_{i=1}^{N} ( \hat{y}_i - y_i )^2 }
\label{eq:rmse}
\end{equation}

where \(y_i\) represents the reference concentration, \(\bar{y}\) denotes the mean reference concentration, and \(\hat{y}_i\) indicates the model-predicted concentration for the \(i\)-th sample. \(N\) corresponds to the number of samples in the evaluated set.

\subsection{Statistical Analysis}

Rigorous statistical comparison of algorithm performance employed non-parametric methods to avoid distributional assumptions. Initial global comparison utilized the Kruskal–Wallis test on validation RMSE values obtained from multiple independent runs. Upon detecting significant global differences, post-hoc pairwise comparisons were conducted using Mann–Whitney U tests with Bonferroni correction for multiple testing. Statistical significance was defined as \(p < 0.05\) after correction.

\section{Results and Discussion}

\subsection{Comparative Algorithm Performance}

Table I presents the GPR model performance metrics for each sample selection algorithm using an 80\% training / 20\% validation split, with random splitting included as a baseline comparator. The Kennard–Stone algorithm demonstrated superior predictive accuracy, yielding the lowest RMSEP (\(1.74 \times 10^{-6}\)) and highest validation \(R^2\) (0.999992). The Honigs algorithm exhibited comparable performance (\(R^2 = 0.999996\), RMSEP \(= 1.76 \times 10^{-6}\)), while Naes and Duplex algorithms produced notably higher validation errors. Random splitting yielded the poorest predictive performance, substantiating the value of rational sample selection methodologies.

\begin{table}[H]
\centering
\caption{Comparative Performance Metrics for Sample Selection Algorithms (80\% Training Proportion)}
\label{tab:performance}
\begin{tabular}{lcccc}
\toprule
\multirow{2}{*}{Algorithm} & \multicolumn{2}{c}{Calibration} & \multicolumn{2}{c}{Validation} \\
\cmidrule(lr){2-3} \cmidrule(lr){4-5}
 & \(R^2\) & RMSE & \(R^2\) & RMSEP \\
\midrule
Kennard–Stone & 0.9999997 & \(2.86 \times 10^{-7}\) & 0.999992 & \(1.74 \times 10^{-6}\) \\
Honigs & 0.9999997 & \(2.61 \times 10^{-7}\) & 0.999996 & \(1.76 \times 10^{-6}\) \\
Naes & 1.0000000 & \(1.45 \times 10^{-8}\) & 0.999984 & \(2.16 \times 10^{-6}\) \\
Random & 0.9999915 & \(2.11 \times 10^{-6}\) & 0.999966 & \(2.30 \times 10^{-6}\) \\
Duplex & 1.0000000 & \(1.65 \times 10^{-8}\) & 0.999932 & \(2.81 \times 10^{-6}\) \\
\bottomrule
\end{tabular}
\end{table}

The elevated calibration performance coupled with degraded validation accuracy observed for Naes and Duplex algorithms suggests potential overfitting, wherein these methods select calibration samples that inadequately represent validation set variability despite excellent calibration statistics. This phenomenon underscores the importance of validation performance as the primary criterion for algorithm selection, as calibration metrics alone may prove misleading.

\subsection{Statistical Validation of Performance Differences}

The Kruskal–Wallis test applied to validation RMSE values revealed highly significant differences among algorithms (\(\chi^2 = 54.32\), \(p = 8.77 \times 10^{-9}\)), rejecting the null hypothesis of equivalent performance. Post-hoc pairwise comparisons employing Mann–Whitney U tests with Bonferroni correction (Table II) elucidated the specific nature of these differences.

\begin{table}[H]
\centering
\caption{Post-Hoc Pairwise Comparisons with Bonferroni Correction}
\label{tab:posthoc}
\begin{tabular}{lccc}
\toprule
Comparison & \(Z\) & Unadjusted \(p\) & Adjusted \(p\) \\
\midrule
Duplex – Honigs & 5.399 & \(6.69 \times 10^{-8}\) & \(6.69 \times 10^{-7}\) \\
Duplex – Kennard–Stone & 5.307 & \(1.11 \times 10^{-7}\) & \(1.11 \times 10^{-6}\) \\
Honigs – Kennard–Stone & -0.092 & 0.927 & 1.000 \\
Duplex – Naes & 2.899 & \(3.74 \times 10^{-3}\) & 0.037 \\
Honigs – Naes & -2.500 & 0.012 & 0.124 \\
Kennard–Stone – Naes & -2.408 & 0.016 & 0.160 \\
Duplex – Random & 1.733 & 0.083 & 0.830 \\
Honigs – Random & -3.666 & \(2.46 \times 10^{-4}\) & \(2.46 \times 10^{-3}\) \\
Kennard–Stone – Random & -3.574 & \(3.51 \times 10^{-4}\) & \(3.51 \times 10^{-3}\) \\
Naes – Random & -1.166 & 0.244 & 1.000 \\
\bottomrule
\end{tabular}
\end{table}

The post-hoc analysis reveals two principal findings. First, both Kennard–Stone and Honigs algorithms significantly outperformed Duplex, Naes, and Random methods (\(p < 0.05\) after correction). Second, and notably, no statistically significant difference emerged between Kennard–Stone and Honigs algorithms (\(p = 1.000\) after correction), indicating their statistical equivalence for this dataset despite numerical differences in point estimates. Figure 1 visually corroborates these findings, illustrating the RMSE distribution for each algorithm.

\begin{figure}[H]
\centering
\includegraphics[width=0.7\textwidth]{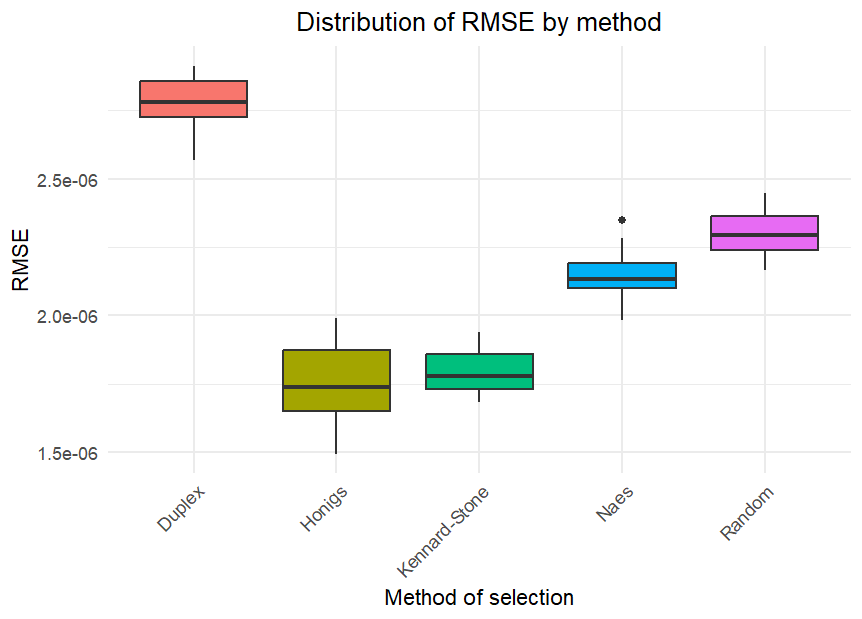}
\caption{Distribution of validation RMSE values by sample selection algorithm, demonstrating the superior central tendency and reduced dispersion for Kennard–Stone and Honigs methods.}
\label{fig:rmse_dist}
\end{figure}

\subsection{Influence of Training Set Size}

Systematic variation of training set proportions from 60\% to 90\% enabled investigation of size effects on model performance for the two top-performing algorithms (Kennard–Stone and Honigs). Table III presents the complete results, while Figure 2 illustrates the observed trends.

\begin{table}[H]
\centering
\caption{Training Set Size Effects on Model Performance}
\label{tab:size_effect}
\begin{tabular}{llcccc}
\toprule
Algorithm & Training (\%) & \multicolumn{2}{c}{Calibration} & \multicolumn{2}{c}{Validation} \\
\cmidrule(lr){3-4} \cmidrule(lr){5-6}
 & & \(R^2\) & RMSE & \(R^2\) & RMSEP \\
\midrule
\multirow{6}{*}{Kennard–Stone}
 & 90 & 0.9999997 & \(2.55 \times 10^{-7}\) & 0.9999985 & \(1.08 \times 10^{-6}\) \\
 & 80 & 0.9999997 & \(2.86 \times 10^{-7}\) & 0.9999921 & \(1.74 \times 10^{-6}\) \\
 & 75 & 0.9999996 & \(3.19 \times 10^{-7}\) & 0.9999879 & \(2.09 \times 10^{-6}\) \\
 & 70 & 0.9999996 & \(3.36 \times 10^{-7}\) & 0.9999619 & \(2.60 \times 10^{-6}\) \\
 & 65 & 0.9999996 & \(3.54 \times 10^{-7}\) & 0.9999276 & \(3.45 \times 10^{-6}\) \\
 & 60 & 0.9999995 & \(4.08 \times 10^{-7}\) & 0.9998787 & \(4.38 \times 10^{-6}\) \\
\midrule
\multirow{6}{*}{Honigs}
 & 90 & 0.9999997 & \(2.53 \times 10^{-7}\) & 0.9999991 & \(1.49 \times 10^{-6}\) \\
 & 80 & 0.9999997 & \(2.61 \times 10^{-7}\) & 0.9999957 & \(1.76 \times 10^{-6}\) \\
 & 75 & 0.9999996 & \(2.97 \times 10^{-7}\) & 0.9999908 & \(2.14 \times 10^{-6}\) \\
 & 70 & 0.9999996 & \(3.06 \times 10^{-7}\) & 0.9999823 & \(2.50 \times 10^{-6}\) \\
 & 65 & 0.9999996 & \(3.25 \times 10^{-7}\) & 0.9999699 & \(2.93 \times 10^{-6}\) \\
 & 60 & 0.9999995 & \(3.66 \times 10^{-7}\) & 0.9999078 & \(4.99 \times 10^{-6}\) \\
\bottomrule
\end{tabular}
\end{table}

The observed monotonic relationship between training set size and predictive accuracy aligns with theoretical expectations from statistical learning theory. As calibration sets encompass greater sample variability, model generalizability proportionally improves. Notably, both algorithms exhibited asymptotic performance approaching the 90\% training proportion, suggesting diminishing returns beyond this threshold.

\begin{figure}[H]
\centering
\includegraphics[width=0.45\textwidth]{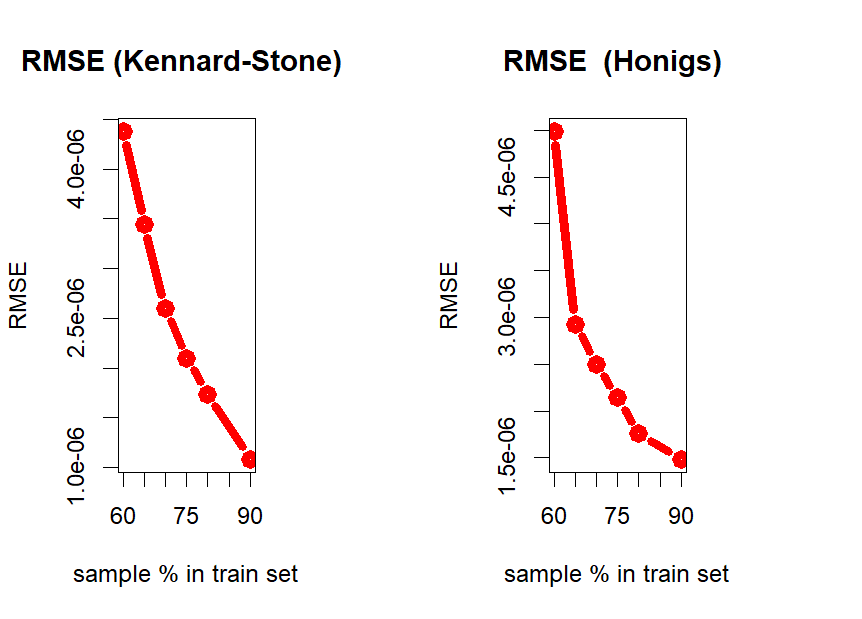}
\includegraphics[width=0.45\textwidth]{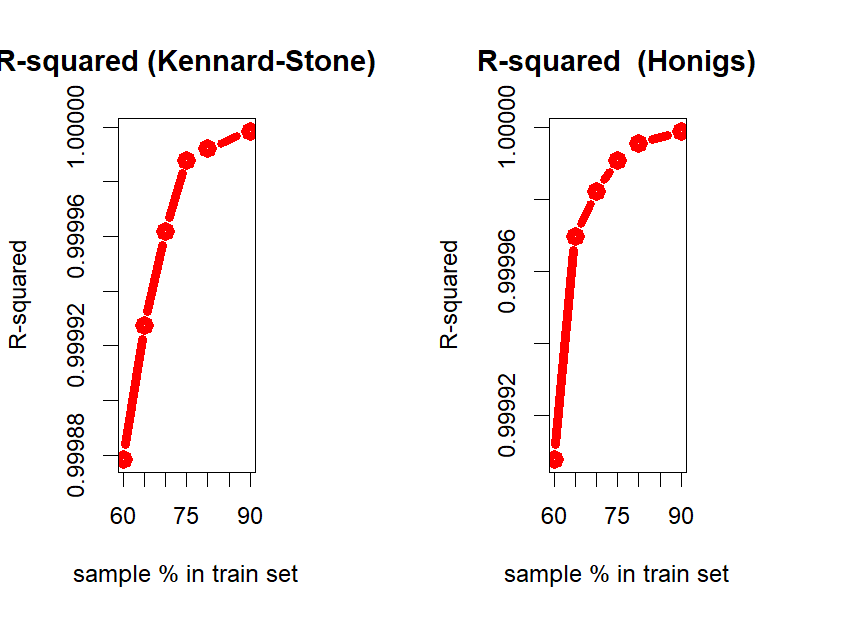}
\caption{Relationship between training set proportion and validation performance metrics: (a) RMSEP demonstrating monotonic decrease with increasing training size; (b) \(R^2\) exhibiting corresponding improvement. Red traces represent Kennard–Stone algorithm; green traces represent Honigs algorithm.}
\label{fig:size_trend}
\end{figure}

Figure 3 provides direct algorithm comparison across all training proportions, visually confirming the statistical equivalence established through hypothesis testing. The near-coincident performance trajectories substantiate the conclusion that both algorithms represent viable options for this application domain.

\begin{figure}[H]
\centering
\includegraphics[width=0.45\textwidth]{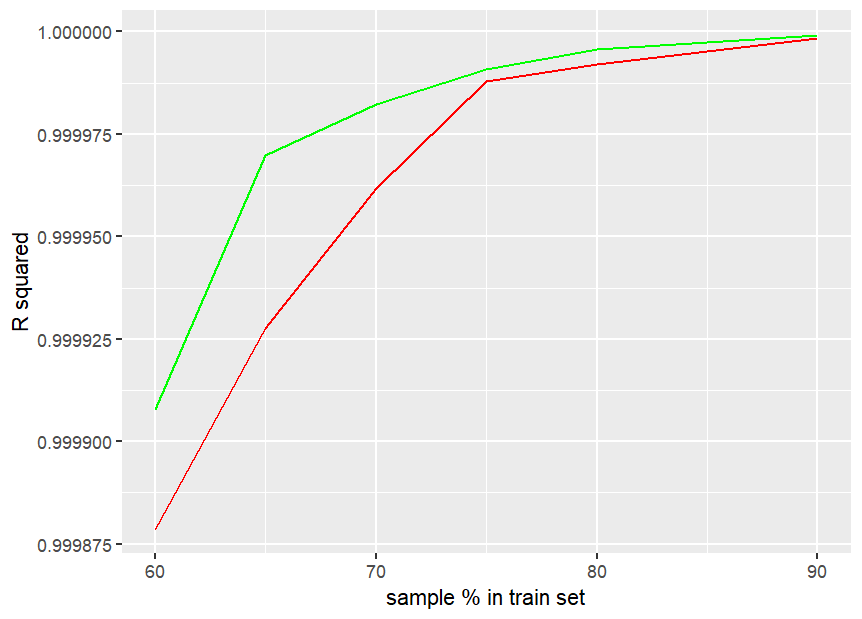}
\includegraphics[width=0.45\textwidth]{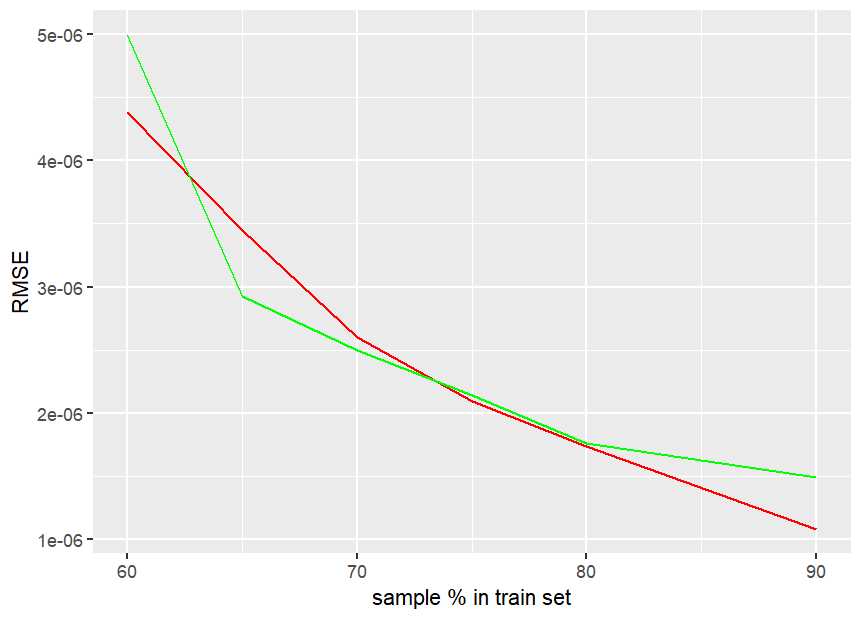}
\caption{Direct comparison of Kennard–Stone (red) and Honigs (green) algorithms across training set proportions: (a) validation \(R^2\); (b) validation RMSEP. The overlapping trajectories confirm statistical equivalence.}
\label{fig:comparison}
\end{figure}

\subsection{Distance Metric Effects}

Investigation of Euclidean versus Mahalanobis distance metrics for Kennard–Stone and Honigs algorithms revealed differential effects (Table IV). For the Kennard–Stone algorithm, Mahalanobis distance yielded superior validation performance (RMSEP \(1.74 \times 10^{-6}\)) compared to Euclidean distance (RMSEP \(1.84 \times 10^{-6}\)). This finding accords with theoretical considerations: Mahalanobis distance accounts for inter-wavelength correlations inherent in NIR spectral data, providing more meaningful dissimilarity measures in high-dimensional correlated spaces. For the Honigs algorithm, both metrics performed comparably, with Euclidean distance showing a marginal advantage.

\begin{table}[H]
\centering
\caption{Distance Metric Effects on Algorithm Performance (80\% Training)}
\label{tab:distance}
\begin{tabular}{llcccc}
\toprule
Algorithm & Distance Metric & \multicolumn{2}{c}{Calibration} & \multicolumn{2}{c}{Validation} \\
\cmidrule(lr){3-4} \cmidrule(lr){5-6}
 & & \(R^2\) & RMSE & \(R^2\) & RMSEP \\
\midrule
\multirow{2}{*}{Kennard–Stone}
 & Euclidean & 0.9999997 & \(2.55 \times 10^{-7}\) & 0.999978 & \(1.84 \times 10^{-6}\) \\
 & Mahalanobis & 0.9999997 & \(2.86 \times 10^{-7}\) & 0.999992 & \(1.74 \times 10^{-6}\) \\
\midrule
\multirow{2}{*}{Honigs}
 & Euclidean & 0.9999997 & \(2.61 \times 10^{-7}\) & 0.999996 & \(1.76 \times 10^{-6}\) \\
 & Mahalanobis & 0.9999997 & \(2.84 \times 10^{-7}\) & 0.999987 & \(1.52 \times 10^{-6}\) \\
\bottomrule
\end{tabular}
\end{table}

\subsection{Methodological Implications}

The findings presented herein carry several implications for chemometric practice in pharmaceutical NIR analysis. First, the demonstrated superiority of rational selection algorithms over random splitting mandates incorporation of systematic sample selection protocols in method development workflows. Second, the statistical equivalence of Kennard–Stone and Honigs algorithms provides practitioners with flexibility in algorithm selection based on computational considerations or implementation convenience. Third, the demonstrated importance of distance metric selection—particularly the advantage of Mahalanobis distance for Kennard–Stone—highlights the need for careful attention to this methodological detail.

The observed overfitting tendency for Naes and Duplex algorithms warrants particular attention. While both methods produced excellent calibration statistics, their degraded validation performance suggests inadequate representation of validation set variability. This finding reinforces the principle that calibration metrics alone provide insufficient evidence of model quality and underscores the necessity of external validation.

\section{Conclusions}

This investigation provides a systematic, statistically rigorous comparison of four sample selection algorithms for NIR spectroscopic calibration in pharmaceutical analysis. The principal findings may be summarized as follows:

1. Rational sample selection algorithms significantly outperform random partitioning, with Kennard–Stone and Honigs methods demonstrating superior predictive accuracy for paracetamol quantification.

2. Statistical analysis employing non-parametric methods with multiple comparison corrections reveals equivalence between Kennard–Stone and Honigs algorithms, while confirming their superiority over Duplex, Naes, and random approaches.

3. Training set size exhibits a monotonic positive relationship with model performance, with optimal results obtained at 80–90\% training proportions for this dataset.

4. Distance metric selection influences algorithm performance: Mahalanobis distance proves advantageous for Kennard–Stone, while Honigs performs comparably with either metric.

5. The observed overfitting in Naes and Duplex algorithms underscores the critical importance of validation-based performance assessment in chemometric model development.

For practitioners developing NIR-based pharmaceutical assays, the Kennard–Stone algorithm employing Mahalanobis distance represents a well-validated, statistically supported approach to sample set partitioning. Future investigations should extend these findings to diverse pharmaceutical product classes and explore advanced optimization techniques such as genetic algorithm-based sample selection for further performance enhancement.

\section*{Acknowledgments}

The authors express their gratitude to the University of Sciences, Techniques and Technologies of Bamako and the Centre of Artificial Intelligence and Robotics of Mali (CIAR-Mali) for institutional support and research facilities.

\section*{Declaration of Conflicting Interests}

The authors declare no potential conflicts of interest with respect to the research, authorship, or publication of this article.

\section*{Funding}

This research received no specific grant from any funding agency in the public, commercial, or not-for-profit sectors.

\section*{Supplemental Material}

Supplemental materials, including R code for all analyses and preprocessing operations, are available from the corresponding author upon reasonable request.

\end{document}